\documentclass[11pt,a4paper]{article}

\usepackage{amssymb}
\usepackage{graphicx}

\newcommand{\be}{\begin{equation}}
\newcommand{\ee}{\end{equation}} 
\newcommand{\bea}{\be \begin{array}{rcl}}
\newcommand{\eea}{\end{array}\ee}
\newcommand{\ba}{\begin{array}}
\newcommand{\ea}{\end{array}}
\newcommand{\ra}{\rightarrow}
\newcommand{\del}{\partial}
\newcommand{\Sp}{\;\;\;\;}

\renewcommand{\Re}{{\rm Re}}
\renewcommand{\Im}{{\rm Im}}

\newcommand{\captionW}[1]{\hspace*{1.3cm}\parbox{13.9cm}{\caption{\footnotesize #1}} }

\newcommand{\half}{{\frac{1}{2}}}

\newcommand{\bra}[1]{\langle{#1}|}
\newcommand{\ket}[1]{|{#1}\rangle}

\newcommand{\dg}{\dagger}

\newcommand{\hU}{\hat{U}}
\newcommand{\hP}{\hat{P}}
\newcommand{\hS}{\hat{S}}
\newcommand{\hX}{\hat{X}}

\newcommand{\Pb}{\overline{P}}

\newcommand{\al}{ \alpha}
\newcommand{\bt}{\beta}
\newcommand{\gm}{\gamma}
\newcommand{\dl}{\delta}
\newcommand{\eps}{\epsilon}
\newcommand{\sg}{\sigma}
\newcommand{\lm}{\lambda}

\title{Spin and  Contextuality in Extended de Broglie-Bohm-Bell Quantum Mechanics}
\author{Jeroen C. Vink  \thanks{\parbox{13.5cm}{
                            Shell Global Solutions International B.V.,
                               Gasweg 31, 1031 HW Amsterdam, The Netherlands.
		      Email: Jeroen.Vink@Shell.com} } }

\begin{document}


\maketitle

\abstract{ This paper introduces an extension of the de Broglie-Bohm-Bell formulation of quantum mechanics,
which includes intrinsic particle degrees of freedom, such as spin, as elements of reality. To evade constraints
from the Kochen-Specker theorem the discrete spin values refer to a specific basis -- i.e., a single
spin vector orientation for each particle; 
these spin orientations are, however, not predetermined, but dynamic and guided by the (reduced, spin-only) wave 
function of the system, which is conditional on the realized location values of the particles. In this way, the
unavoidable contextuality of spin  is provided by the wave function and its realized particle configuration, whereas
spin is still expressed as a local property of the individual particles.
This formulation, which furthermore features a rigorous discrete-time stochastic dynamics, allows for numerical
simulations of  particle systems with entangled spin, such as Bohm's version of the EPR experiment.
}



\section{Introduction} \label{sect1}
One of the ongoing debates around the foundations of quantum mechanics is  the question to what extent  concepts
and properties of classical physics continue to apply to quantum physics.
The results of measurement processes  in classical physics simply reflect  properties of one or more 
elements of reality that exist independently of observers or measurements. 
These elements of reality, particles or electromagnetic fields for example, 
have their time dependent values regardless of this measurement or observation ``context'',
 and their evolution is only influenced by their direct surroundings. This non-contextual nature of reality has never 
been at issue within classical physics, unlike the desire for local causality, which, however, had become firmly
established with the development of Maxwell's theory of electromagnetism and Einstein's relativity theories. 
It replaced the unpalatable non-local action at a distance present in Newton's and Coulomb's theories, where a 
change in location of a massive or charged body had an  instantaneous impact on bodies that were arbitrary far away.

Quantum mechanics in the Copenhagen interpretation denies the existence of a microscopic reality in which particles
and fields have properties that are independent of observations or measurements. 
This notion of reality without (counterfactual) definiteness was challenged in the famous
paper by Einstein, Rosen and Podolsky \cite{EPR35}, who pitted locality of the quantum theory against its completeness.
It would require non-local action at a distance, they argued, to avoid the conclusion that measurement results 
exist independently from the act of measuring and hence a quantum reality with only an evolving or collapsing 
wave function cannot be complete.
Somewhat ironically, in a reversal of the EPR argument, Bell showed \cite{Bell64} that these more complete, i.e., 
hidden variable versions of quantum mechanics are necessarily in conflict with local causality if they are to 
reproduce the results computed with the well-established rules of quantum mechanics. Therefore it follows
\cite{Norsen05,BellCh10} that the combined EPR and Bell arguments imply that quantum mechanics, with or 
without hidden variables, does not support the classical notion of locality.
Furthermore, the counterfactual nature of reality was challenge by the almost equally famous theorem by Kochen and 
Specker \cite{KochenSpecker}, which forbids the 
existence of a non-contextual value map. I.e., in a quantum system, it is not possible to simultaneously assign 
definite values to all Hermitian operators in the system's Hilbert space (while maintaining the usual functional relationships between these values).
Since measurements single out a specific set of (mutually commuting) Hermitian operators, 
it appears that values of microscopic elements of reality, if these exists, must somehow be dependent 
on a macroscopic measurement context.

This picture is different in Bohm's interpretation of quantum mechanics \cite{deBroglie56,Bohm52}, 
where he showed that 
the location of particles can in fact exist as (counterfactually real and non-contextual) elements of reality. Restoring
such an observer-independent ontology, which  lends itself to numerical simulations of a microscopic quantum
world, has subsequently provided valuable  insights into the still distinctly non-classical features of quantum
physics - for example in the field of quantum chemistry \cite{QuantChem}
and quantum cosmology \cite{PintoNeto13}.

However, one may wonder why, in Bohm's interpretation, only particle positions (in non-relativistic quatum mechanics) 
have this  factual existence. 
Additional, intrinsic degrees of freedom that are also represented in the system's Hilbert space,
such as particle spin are absent in this quantum world and must remain  properties that only manifest
themselves indirectly through their impact on particle trajectories
 \cite{BohmHileyCh10,Norsen14,BricmontGoldsteinHemmick19a}.
Consensus, at least among adherents of Bohm-style interpretations, is that  it  is both sufficient and natural 
for ``position'' (in contrast to for example ``momentum'') to be the only observable with a factual existence. 
This is typically substantiated by stating that all observations and experimental results are 
ultimately realized in terms of locations of pointers or dots on a screen or paper. 

But one could challenge 
this assertion: if, on a dark night, I see a very faint star, because just a few photons hit the retina in my eye, do I 
then observe the location of these photons or is it their momentum that triggers a signal that seeds my awareness 
of the star's faint light from which I infer its location on the firmament? 
Or, more importantly, I would argue that my awareness of a measurement result
does not only involve a pointer in a specific location, but ultimately consists of a specific, realized state of the visual 
cortex in my brain. It is not clear that the realized values of {\it only} the positions of the particles in my brain 
is sufficient to express the mental state of seeing pointer positions or a faint star. 
Possibly, also particle spin or  other internal particle
degrees of freedom must assume realized values in order to represent the richness of our conscious minds - i.e.,
perhaps also these quantum features must be present as elements of reality.

An additional, possibly more compelling reason to extend the minimal Bohm interpretation presents itself when 
considering situations where the omission of internal degrees of freedom in the set of elements of reality 
leads to particle trajectories with remarkable features
-- so much so that they have been characterized as ``surreal'' \cite{Englert93,Scully98}. Explaining and justifying
these unexpected features of Bohmian trajectories requires careful analysis and invocation of the 
non-local contextuality of particle properties other than location \cite{Durr93,Dewdney93,Hiley00,Barrett00}. 
Including internal particle degrees of freedom into the set of elements of reality,
on the other hand, results in particle trajectories that naturally conform with one's intuitive expectations, as is
further elaborated on in Appendix \ref{Ap_A}.

In this paper I will, therefore, explore an extension of the de Broglie-Bohm-Bell approach (dubbed eBBB) which includes 
particle spin. 
Because position is indeed such a natural and obvious candidate to be part of reality and since there 
are additional supporting arguments \cite{Wiseman07} to single-out location as 
a preferred observable, I will not further challenge this assumption. 
As was shown in ref.~\cite{Vink93},  it is straightforward to apply Bell's (re)formulation of Bohmian dynamics
\cite{BellCh19} to generate value trajectories for spin provided one chooses a specific spin representation 
(i.e., a preferred basis). 
However, there is no natural, fixed choice for such a preferred spin representation, since it depends on the 
specific (experimental) context in which orientation the spin value is measured. This problem can be overcome through 
an additional rule for the stochastic dynamics that  automatically selects the most natural representation. 
In the present paper this is achieved by dynamically adjusting the spin representation through guidance
of the wave function and conditionally on the realized values of the system's location variables.\footnote{This approach 
combines elements from the work by Dewdney et al.~\cite{Dewdney87}, who constructed guidance equations for the 
Euler angles of a system with two spinning particles, and the Bell-type approach from ref.~\cite{Vink93}.  
In this early work, the approach to represent quantum spin as a directional (angular momentum-like) vector 
for each particle was abandoned, because such a collection of 3D spin vectors could not accommodate the 
exponential growth of the dimension of the spin state space when the number of particles increases 
(cf.~the discussion in sections 10.2 and 10.3 of ref.~\cite{BohmHileyCh10})} 
This shows explicitly that the required context dependence 
need not be provided by an (external) experiment, but can be supplied by the dynamically evolving system itself.

It is a key benefits of the eBBB approach that it allows for computer
simulations of this enriched microscopic quantum world. The ``simulatability'' of this formulation is further strengthened
in this paper by a rigorous extension of the stochastic  dynamics to discrete time. 
This not only allows performing numerical simulations of quantum systems without the need for further 
approximations to solve the stochastic trajectory evolution along with the Schr\"odinger equation,
it also simplifies the notion of time and suggests a natural definition for the discrete time steps.
It would be interesting to see if similarly detailed and credible simulations of quantum systems that can be
performed with the eBBB formulation, 
for example of Bohm's version of the EPR experiment, can be achieved using  other formulations of 
quantum mechanics that contend not to rely on external observers, e.g.,  Everett's many-worlds 
interpretation \cite{Everett57},  the different versions of consistent histories formulations
 \cite{Griffiths84,Omnes88,Hartle90} and the spontaneous collapse interpretation \cite{GWR86} 
(see e.g.~refs.~\cite{Goldstein98,Barrett19} for an overview of these approaches).

The remainder of the paper is organized as follows.  
The next section describes the extensions of the eBBB formulation after which
section \ref{sect3} focuses on simulations of two types of quantum spin systems. First, a simulation of
Larmor precession of  an entangled pair of spin-two particles is described, followed by
a description of the set-up and results of a simulation of a simplified version of the EPRB experiment
(Bohm's modified version of the EPR experiment).  
The closing section contains a brief summary and discussion. An elaboration of the different behavior of
trajectories in the causal Bohm and eBBB formulations, and technical details of the numerical EPRB 
experiment are relegated to two appendices.

\section{Extended de Broglie-Bohm-Bell Formulation} \label{sect2}
The extended de Broglie-Bohm-Bell (eBBB) formulation described below introduces three  improvements over the  
approach of refs.~\cite{BellCh19} and \cite{Vink93}: 
First, the continuous time formulation of  the stochastic dynamics is replaced by a discrete time formulation; 
second, it is shown that the beable dynamics can be expressed using a basis of the Hilbert space that changes in time;
third, a special dynamics is proposed  for the basis (or representation) of the internal degrees of freedom of 
the particles (such as spin). With this dynamics, the representation for particle spin will, conditionally on the 
attained location realization, 
adjust to a representation that is most appropriate to express the physical, for example measured, state of the system.

\subsection{The eBBB Formulation in Discrete Time} \label{sect2sub1}
The formulation of discrete time eBBB dynamics starts with the fundamental evolution equation for a quantum state,
\be  
    \ket{\psi}^{t+1} = \hat{U}\ket{\psi}^t.           \label{Eq_1}
\ee
Here, $\ket{\psi}^t$ represents the  quantum state at time step $t$, which is propagated to the state at the next time 
step $t+1$ through the action of the unitary evolution operator $\hat{U}$. This operator follows from the  Hamiltonian operator $\hat{H}$, as
\be
  \hat{U} = e^{-i \hat{H}\eps }.          \label{Eq_3}
\ee
The Hamiltonian may, and the time step size $\eps$ will depend on the index $t$, which labels are suppressed to
avoid cluttering the notation too much. The relevance of a variable time step size will become clear shortly.

As in  ref.~\cite{Vink93}, the system's Hilbert space is assumed to be finite with dimension $N$, such that operators can 
be represented as Hermitian matrices and wave functions as vectors.\footnote{This assumes that space is both finite 
and discrete, consisting of discrete points; even though quantum states are technically represented as vectors, 
I will often refer to them as wave functions.} 
The Schr\"odinger equation (\ref{Eq_1}) and an arbitrary (Hermitian or unitary) 
operator $\hat{O}$ in  the $n$-representation  can  be written as
\be
     \psi_n^{t+1}= \sum_{m=1}^N  U_{n,m} \psi_m^t , \Sp \psi_n^t=\bra{n}\psi\rangle^t,   \label{Eq_2b}
\ee
and
\be
   O_{n,m}=\bra{n}\hat{O}\ket{m}.    \label{Eq_2}
\ee
respectively.

In order to achieve the same result as Bell  obtained using the differential form of the state evolution \cite{BellCh19},
the discrete time evolution of the probabilities  $P_n^t=\psi_n^{*t}\psi_n^t$ should be linked to an anti-symmetric
probability current and then expressed in terms of  transition probabilities between different values of the 
eigenvector  index $n$.
This can be achieved as follows. First, the probabilities for the next time step are written as
\be
   P_n^{t+1} =  \psi_n^{*t+1}\psi_n^{t+1} = \sum_{m}\psi_n^{*t+1}U_{n,m}\psi_{m}^t.   \label{Eq_4}
\ee
Next,  a real-valued matrix $K$ is defined as
\be
 K^t_{n,m} = \Re( \psi_n^{*t+1}U_{n,m}\psi_{m}^t ),   \label{Eq_5}
\ee
which can be used to write
\bea
    P_n^{t+1}  & = & \sum_m K^t_{n,m} = \half\sum_m (K^t_{n,m} + K^t_{m,n}) + \half \sum_m ( K^t_{n,m} - K^t_{m,n} ) \\[2mm]
                        & = & \half\sum_{km}( \psi_k^{*t}U_{kn}^{\dg}U_{n,m}\psi_{m}^t
                                            + \psi_k^{*t}U_{km}^{\dg}U_{m,n}\psi_{n}^t) +\half\sum_m(K^t_{n,m} - K^t_{m,n}) \\[2mm]
                       & = & \half   \psi_n^{*t+1}\psi_n^{t+1} + \half \psi_n^{*t}\psi_n^{t} 
                                            +\half\sum_m  (K^t_{n,m} - K^t_{m,n}).  \label{Eq_7}
\eea
After defining the anti-symmetric probability current matrix $J^t$ as
\be
   J^t_{n,m} =  K^t_{n,m} - K^t_{m,n} = \Re (\psi_n^{*t+1}U_{n,m}\psi_{m}^t 
                                                                  -  \psi_m^{*t+1}U_{m,n}\psi_{n}^t) ,     \label{Eq_8}
\ee
equation (\ref{Eq_7}) turns into a discrete-time continuity equation
\be
          P_n^{t+1}  =  P_n^t +  \sum_m J^t_{n,m}.  \label{Eq_9}
\ee
Following  similar steps as in ref.~\cite{BellCh19}, this continuity equation can be worked into the desired
master equation,
\bea
 P_n^{t+1} & = & P_n^t + \sum_m \left( \theta(J^t_{n,m}) J^t_{n,m}   
                                                           - \theta(J^t_{m,n}) J^t_{m,n} \right), \\[2mm]
                    & = & P_n^t + \sum_m( T^t_{n,m} P_m^t  - T^t_{m,n} P_n^t),             \label{Eq_9b}
\eea
where $\theta(x)$ is the Heaviside step function and transition probabilities are defined as
\be
   T^t_{n,m} = \theta(J^t_{n,m}) J^t_{n,m} / P^t_m =
                              {\rm max}\left(0, \Re(\psi^{*t+1}_n U_{n,m} \psi^t_m )- (n \leftrightarrow m)  \right)/ P^t_m.
                            \label{Eq_10}
\ee

It should be noted that this  equation for the time evolution of $P_n$ is an exact result but only makes sense as 
a master equation if the transition probabilities satisfy the consistency condition
\be
  T^t_{m,m} \equiv 1 - \sum_{n \ne m}T^t_{n,m} \ge 0 \Sp \forall m,    \label{Eq_11}
\ee
which  automatically implies that
\be
 T^t_{n,m} \le 1 \Sp \forall n,m.    \label{Eq_12}
\ee
Eq.~(\ref{Eq_10}) suggests that condition (\ref{Eq_11}) can  be violated when for one or more $m$, 
the probability  $P^t_m\ra 0$.
A similar consistency condition on transition probabilities in Bell's result is formally  satisfied owing to the 
infinitesimal value of the time step size in his continuous time formulation. This suggests to explore if the 
consistency conditions for the discrete time formulation will be met for sufficiently small time steps. 

For small time step size $\eps$ the evolution matrix can be approximated as
\be 
  U_{n,m} = \dl_{n,m} - iH_{n,m} \eps + O(\eps^2).  \label{Eq_13}
\ee
Substituting this in Eq.~(\ref{Eq_10}) and collecting terms up to order $\eps$, one can verify that the result 
obtained by Bell \cite{BellCh19} is recovered
\be
    T^t_{n,m} = \tilde{T}^t_{n,m} \eps + O(\eps^2),
\ee
where the transition probability {\it rate} matrix $\tilde{T}^t$  is defined as
\be 
  \tilde{T}^t_{n,m} = {\rm max}\left(0,-2\Im(\psi^{*t}_n H_{n,m}\psi^t_m) \right) / P^t_m .
\ee
Writing $\psi^t_m=R^t_m e^{iS^t_m}$, it follows that
\be
    T^t_{n,m} = {\rm max}\left( 0,-2\Im(H_{n,m}e^{i(S^t_m-S^t_n)} )\right)(\eps R^t_n/R^t_m)
                                                                                                            + O(\eps^2).  \label{Eq_Teps}
\ee
This shows that $1-T^t_{m,m}\sim \sum_{n\ne m} c^t_{n,m}\eps/R^t_m$ for $\eps\ra 0$, where the $c_{n,m}$ are
positive numbers. Since the quantum system is finite and the state with index $m$ has been realized,
the value of $P_m$ is non zero (i.e.~$R^t_m > \dl^t$ with $\dl^t > 0$) and the value of 
$ \sum_n c^t_{n,m}\equiv M^t$ is finite. 
This in turn implies that the constraint (\ref{Eq_11}) can be satisfied with a finite time step size
that is chosen sufficiently small, $\eps < \dl^t/M^t$.

In fact, when this eBBB approach is applied to an isolated quantum system, such as the Universe \cite{Vink92},
the value of $\eps$ is unobservable since there is no external clock. 
Therefore, a natural choice for a discrete, time index dependent time step
value is that it is always maximally large, such that Eq.~(\ref{Eq_11}) just holds.
For a large system, the dscrete time steps $\eps^t$ will then  naturally be very small, since the bounding 
values $\dl^t$ and $1/M^t$ will be very small.
In practical simulations to compute an ensemble of trajectories \cite{Vink93,Vink18}, 
it has been sufficient to choose a reasonably small default value for $\eps$ and (temporarily) reduce this value
when a state $m$ with a very small $P_m$ would occur; such a state will then with high probability jump to 
one that has a much larger probability, which  allows restoring the original time step size.

\subsection{Trajectories with a  Dynamically Changing Beable Representation}   \label{sect2sub2}
 In the discussion above, a fixed $n$-representation 
was used to obtain a prescription for generating trajectories for the eigenvalue index values in this representation. 
However, it is possible to generalize this approach to allow using a representation that dynamically
changes during the evolution of the trajectories. This is an interesting feature because it can be used to address the
Preferred Basis problem \cite{Barrett05,deRonde16}. 
In particular, the present paper will show how the trajectory dynamics can be extended
to generated an optimal spin basis along with  spin (and location) values for the particles.

Starting from the Schr\"odinger equation in the $n$-representation, Eq.~(\ref{Eq_2b}), a change to a different
$\al$-representation at time $t$ involves transforming $\psi$ and $U$ according to,
\be
   \psi^{V^t}_{\al} = \sum_n V^t_{\al,n}\psi_n    \label{VonPsi}
\ee
and
\be
   U^{V^{t+1}V^t}_{\al,\bt} = \sum_{nm}V^{t+1}_{\al,n}U_{n,m}V^{t\dg}_{m,\bt}   \label{VonU}
\ee
respectively, with $V^t_{\al,n} = \bra{\al}n\rangle$ the unitary transformation from the $n$- to the 
$\al$-representation.
A different change in representation can be applied at every time step, such that the  index values of
state vectors $\psi_{\al}^t$ and $\psi^{t+1}_{\bt}$ refer to different representations $V^t$ and $V^{t+1}$:
\be
  \psi^{V^{t+1} t+1}_{\bt} = \sum_{\al}U^{V^{t+1}V^t}_{\bt,\al}\psi^{V^t t}_{\al}. \label{Vtrans}
\ee
Hence, the value of the eigenvalue index is no longer sufficient to describe the trajectory state; 
also the transformation matrix $V^t$ or another specification of the representation used at each time step 
must be provided and therefore has become  a dynamic element of reality. 

Since the dimensionality of the system's Hilbert space is very large, there obviously is a lot of freedom
to choose basis transformations $V$. 
To make the challenge of finding a credible basis dynamics tractable, I will make three assumptions: 

First,
the basis transformation must be determined by the system's wave function. External and internal degrees
of freedom are handled separately. In this paper, focus will be on basis changes for internal degrees of freedom,
for which the wave function that guides basis changes will be restricted to these internal degrees of freedom.
As will be shown in more detail below, this implies that a different (spin) basis change may be applied for each
different value of the configuration argument in the wave function.

Second,
in a multi-particle system, the combined basis must be a direct product of single-particle bases.
This ensures that the eigenvalue trajectories can be decomposed into independent values, such that
each particle in the system has its own well-defined and localized properties. 

Third, 
the admissible (single particle) basis sets will be restricted to those  that correspond to  operator representations 
that are relevant for, and observed in the macroscopic world (e.g., location, momentum, angular momentum, 
3D spin-orientation, etc.). The optimal basis choice will then be the one in which the (reduced) state vector maximally
aligns with one of the vectors in this basis.

The first assumption to only apply basis transformations to the spin (or other internal) degrees of freedom
can be made explicit by writing the generic eigenvalue index $n$ as a pair $xs$, where $x$ refers
to the locations of the particles and $s$ to their spin. Using this notation, the detailed expression for 
basis changing transition probabilities becomes
\be
T^{V^{yt}V^{xt} t}_{ys',xs} ={\rm max}(0,\Re(\psi^{*V^{yt}(t+1)}_{ys'}
                                       U_{yu',xv'}^{ V^{yt} V^{xt} }\psi^{V^{xt}t}_{xs}) 
                                     -  (xs\leftrightarrow ys')   )/ P^{V^{xt}t}_{xs},   \label{Eq_TVx}
\ee
with
\be
   P^{V^{xt}t}_{xs} = \vert \psi^{V^{xt}t}_{xs}\vert^2.
\ee
The basis transformations $V$ now only work on spin indices, and may depend both on time index $t$ and
configuration index $x$. I.e., the transformed wave functions and evolution matrix are,
\be
   \psi^{V^{xt}t}_{xs} = \sum_{v}  V^{xt}_{s,v} \psi^{t}_{xv}, \Sp 
    U_{ys',xs}^{ V^{yt} V^{xt} }  = \sum_{uv}  V^{yt}_{s',u}U_{yu,xv} V^{xt\dg}_{v,s}.
\ee
Also with these conditional ($x$-dependent) basis transformations the steps in  Eqs.~(\ref{Eq_4}-\ref{Eq_10}) 
can be followed to verify that the transition probabilities (\ref{Eq_TVx}) generate the same time-dependent
probability distributions $P^t_{xs}$ as are produced through the Schr\"odinger dynamics in quantum mechanics.
I.e., if $P^0_{xs}= \vert \psi^{V^{x0}0}_{xs}\vert^2$ at $t=0$, then $P^t_{xs}$ computed using
the transition probabilities (\ref{Eq_TVx}) will be equal to $\vert \psi^{V^{xt}t}_{xs}\vert^2$ for all $t$.

The next section describes  how appropriate spin basis transformations $V^{xt}$ can be computed from  the system's
wave function $\psi^t_x$.

\subsection{A Self-Adjusting Spin Basis}   \label{sect2sub3}
The wave function for $N$ particles with spin can be written in more detail as
\be
   \psi^t_{xs} \equiv \psi^t_{x_1s_1,x_2s_2,\dots,x_Ns_N}
\ee
where the pairs $x_i s_i$  indicate the location (external) and spin (internal) degrees of freedom
of particle $i$.
In accordance with the first assumption described above,  a preferred, dynamically adjusting basis for the spin
subspace of this system will be pursued, which should be derived from the spin-content of the system's wave function.
For a given configuration $x$, the associated conditional spin-only wave function is  
\be
 \tilde{\psi}^{xt}_{s_1,s_2,\dots,s_N} =  \psi^t_{x_1s_1,\dots,x_Ns_N}/
                    (\sum_{s_1,\dots,s_n}\vert\psi^t_{x_1s_1,\dots,x_Ns_N}\vert^2)^{1/2}.     \label{Eq_PsiS}
\ee
The preferred basis in this spin subspace will be the one in which the reduced state vector (\ref{Eq_PsiS})
maximally aligns with one of the basis vectors.

Next, the third assumption is used to limit candidate basis sets to those that represent different orientations of the
spin vector. I.e., the collection of candidate single-spin basis sets are assumed to be 
eigenvector sets of the $z$-component of an arbitrarily rotated spin operator $\vec{\hS}=(\hS^x, \hS^y, \hS^z)^T$.
These rotations can be parameterized with two Euler angles, $\theta$ (the angle between the
rotated $z$-axis and the reference $z$-axis and $\phi$ (the angle of rotation around the $z$-axis).

To obtain   Euler angles for the spin basis for each separate particle, the second assumption is used, which states that
candidate basis vectors of the multi-spin system will be tensor products of the form
\be
     v^{m_1}_{\theta^t_1\phi^t_1}\otimes\dots \otimes v^{m_N}_{\theta^t_N\phi^t_N}.
\ee
Here,  the basis vectors $v^m_{\theta^t \phi^t}$ with $m=-s,\dots,s$ are the eigenvectors of $\hS_z$ 
rotated by (time-dependent) Euler angles $\theta^t$ and $\phi^t$. 

The  optimization that is required to find the basis in which the reduced state vector is maximally aligned with
one of the basis vectors  can be performed in two steps. First, the multi-spin wave function (\ref{Eq_PsiS}) 
is approximated by a product of single spin wave functions, by performing the minimization
\be
   \min_{\psi^{(1)},\dots,\psi^{(N)}}\sum_{s_1\dots s_N=1}^{n_s}
                             \vert\tilde{\psi}^{xt}_{s_1,s_2,...,s_N} - \psi^{(1)}_{s_1}\dots\psi^{(N)}_{s_N}\vert^2
\label{Eq_minPsi}
\ee
subject to $\vert\psi^{(i)}\vert^2=1$ for all $i$. 
Second, these individual wave functions $\psi^{(i)}$ are matched against candidate basis vectors.
 I.e., for all $i$ the optimal rotation angles $\theta^t_i$ and $\phi^t_i$ and best-fitting basis
vector index $m_i$ are determined by minimizing
\be
    \min_{\theta^t_i,\phi^t_i,m_i}\vert\psi^{(i)}  - v^{m_i}_{\theta^t_i \phi^t_i}\vert^2.  \label{Eq_bvFit}
\ee

In general the minimization (\ref{Eq_minPsi}) is difficult to perform analytically, but for $N=2$ it amounts to solving
the two eigenvalue problems
\be
   \sum_{s_2,s_3} (\tilde{\psi}^*_{s_3,s_1}\tilde{\psi}_{s_3,s_2}) f^{\lm}_{s_2} = \lm f^{\lm}_{s_1} \Sp{\rm and}\Sp
   \sum_{s_2,s_3} (\tilde{\psi}_{s_1,s_3}\tilde{\psi}^*_{s_2,s_3}) g^{\mu}_{s_2} = \mu g^{\mu}_{s_1}.
   \label{Eq_Ev}
\ee
The desired wave function factors are the eigenvectors with largest eigenvalue, $\psi^{(1)}=f^{\lm_{max}}$
and $\psi^{(2)}=g^{\mu_{max}}$. In case of degeneracy of this maximum eigenvalue, an additional condition must be
imposed\footnote{A two-fold degeneracy does occur in one of the examples to be discussed below. The ambiguity
can be fixed, for example, by choosing the linear combination that maximally aligns with the constant vector $(1,1,\dots,1)^T$} 
to ensure finding a well-defined and unambiguous result when repeatedly computing 
$\psi^{(1)}$ and $\psi^{(2)}$ for successive time steps. 

To avoid heavy notation, the superscript $xt$ is
not always shown in Eqs.~(\ref{Eq_minPsi}-\ref{Eq_Ev}). It should be noted, however, that the
preferred spin basis in general will be time-dependent and may be different for each different 
particle configuration $x$.

\section{Examples of eBBB Trajectories} \label{sect3}
In this section the above formulation to make  both particle locations and spin values elements of reality is applied 
to two (simplified) quantum systems. The first example focuses on the spin dynamics and shows trajectories
for spin values and  orientation angles of the spin vector. The pure spin system mimics two static but entangled 
spin-two particles in a magnetic field, such that the spin vectors exhibit Larmor precession.
The second example illustrates the combined dynamics of particle locations and spin values in
 a simplified realization of Bohm's version of the EPR experiment.
In this example the quantum system consists of the two entangled particle spins {\it as well as} the  
angles of the Stern-Gerlach magnets that, last-minute, determine the spin orientations of the deflected particles.

\begin{figure}[tth]
\begin{center}
\includegraphics[height=6.5cm]{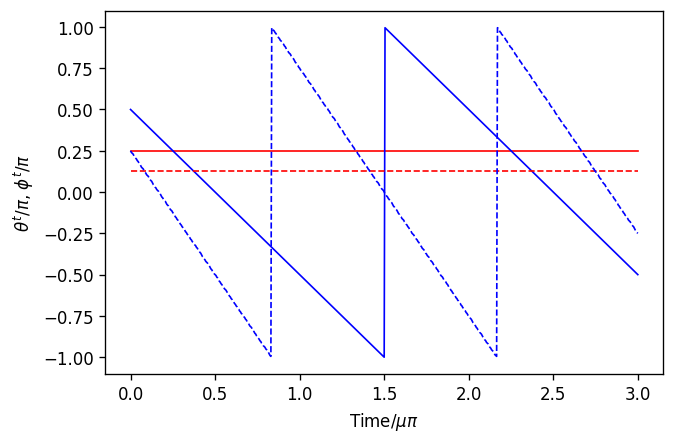}
\end{center}
\captionW{Euler angles $\theta^t$ (horizontal lines) and $\phi^t$ for the system (\ref{Eq_2Psys}) 
with two entangled spin-2 particles in a magnetic field  with interaction strength $\mu_1=\mu$ for particle
one (solid lines) and $\mu_2=3\mu/2$ for particle two (dashed lines);
the initial state has Euler angles $\theta_1=\pi/4$ and $\phi_1=\pi/2$ for particle one and
$\theta_2=\pi/8$ and $\phi_1=\pi/4$ for particle two.
\label{Fig_2SpAngles}}
\end{figure}

\begin{figure}[ttt]
\begin{centering}
\scalebox{0.8}{\includegraphics[height=3.0cm]{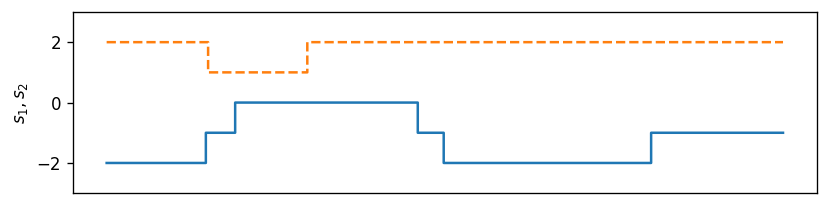}}\\[-2mm]
\scalebox{0.8}{\includegraphics[height=3.0cm]{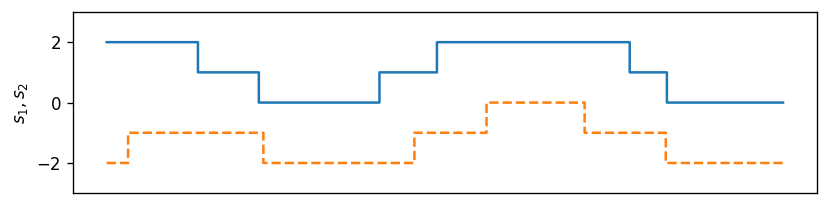}}\\[-2mm]
\scalebox{0.8}{\includegraphics[height=3.0cm]{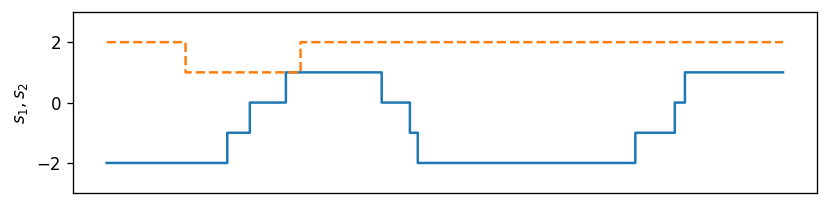}}\\[-2mm]
\end{centering}

\captionW{Three trajectories for spin $s_1$ and $s_2$ for the same system as in Figure~\ref{Fig_2SpAngles}.
The solid and dashed lines are for particle one and two respectively.
\label{Fig_2SpTraj}}
\end{figure}

\begin{figure}[ttt]
\begin{center}
\includegraphics[height=6.5cm]{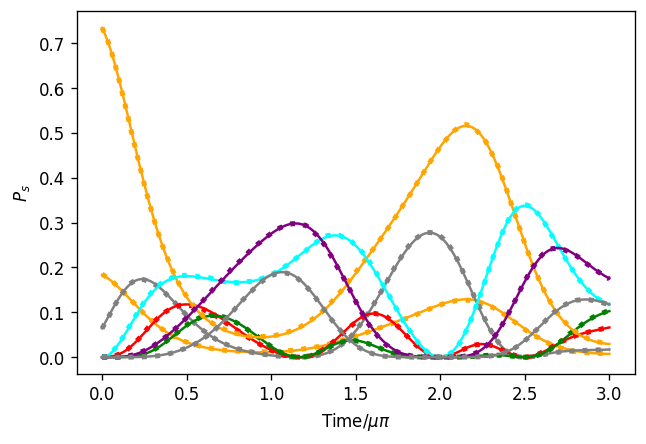}
\end{center}
\captionW{Analytical (solid lines) and ensemble averages (dots) of the time-dependent spin-index values 
for the same system as in Figure~\ref{Fig_2SpAngles}.
The 8 curves show the time-dependent probabilities $P^t_s$ with $s=s_1+2+5(s_2+2)$ and $s_1,s_2\in[-2,2]$,
for which $\max_t{P_s^t}>0.1$. The ensemble size used to compute the averages is 50,000.
\label{Fig_2SpProb}}
\end{figure}

\subsection{Two Entangled Particles with Larmor Precession} \label{sect3sub2}
The Hamiltonian for a particle with spin $s$ in a magnetic field pointing in the $z$-direction can be written as
\be
 \hat{H}_{\mu} = -\mu \hS_z,   \label{Eq_Hmu}
\ee
with $\mu$ representing  the strength of the interaction of the magnetic field and the particle's dipole moment  
and $\hS_z$ the $z$-component of the spin operator in an $n_s = 2s+1$ dimensional Hilbert space. 
The particle is assumed to be sufficiently heavy, such that
the kinetic term in the Hamiltonian can be ignored. To make this example more interesting, the spin 
is taken to be relatively large, such that the dimension $n_s$ of the spin
Hilbert space is  lager than the dimension of the 3D rotation group. 

A system consisting of two such spin-two particles with different magnetic moments, 
has an $n_s\times n_s=25$ dimensional Hilbert space, with Hamiltonian
$\hat{H} = \hat{H}_{\mu_1}+ \hat{H}_{\mu_2}$ and a  wave functions $\psi_{s1,s_2}$.
To  test the robustness of the formulation of section \ref{sect2sub3},  the initial wave function
is chosen to be a non-separable combination of rotated basis vectors, 
\be
      \psi^0 = (v^{m_1}_{\theta_1 \phi_1} \otimes v^{m_2}_{\theta_2 \phi_2} -     \label{Eq_2Psys}
                             2 v^{m_2}_{\theta_1 \phi_1} \otimes v^{m_1}_{\theta_2 \phi_2} )/\sqrt{2},
\ee
with $\theta_1 = \pi/2$ and $\phi_1 = \pi/4$ for particle one and 
$\theta_2 = \pi/4$ and $\phi_2 = \pi/8$ for particle two; the (unrotated) spins are maximally aligned with
the $z$-axis and have opposite sign: $m_1=2$, $m_2=-2$.
Also the magnetic coupling strengths are different: $\mu_1 = \mu$ and $\mu_2 = 3\mu/2$ for particle
one and two respectively.

The results in Figure~\ref{Fig_2SpAngles} confirm that the microscopic system exhibits the expected dynamics 
for the Euler angles: the two particle spins precess with  the frequency expected from the strength of the 
magnetic coupling, with a constant tilt angle of the spin vector.
These tilt angles $\theta_{1,2}$ and start values $\phi^0_{1,2}$ correspond exactly with the
Euler angles of the rotation of the basis vectors that define the initial state.

The formulation of section \ref{sect2sub2} was used to compute an ensemble of trajectories with spin values
in these continuously changing representations. Three of these trajectories are shown in Figure \ref{Fig_2SpTraj},
where the solid and dashed lines are the trajectories for the spin of particle one and two respectively.
Since the evolving state vector of the entangle spins never is exactly aligned with the (rotating) basis vectors,
the spin value can jump away from their initial values. Finally, Figure~\ref{Fig_2SpProb}
shows that the ensemble averages of these generated spin values exactly reproduce the probabilities expected from
the evolving wave function $\psi^{Vt}_s=\sum_{s'}V^t_{s,s'}\psi^t_{s'}$.
 The plot shows eight curves (solid lines) of $P^t_s=\vert\psi^{Vt}_s\vert^2$  for which $\max_t{ P^t_s}>0.1$; 
the dots that overlay these lines are the corresponding values computed from the ensemble of spin trajectories.

\subsection{The Einstein-Podolsky-Rosen-Bohm  Spin Experiment} \label{sect3sub3}
This section describes the set-up and results of simulations of a simplified model for Bohm's version of the 
EPR experiment (EPRB for short). Here the spin dynamics and representation will need to be conditional
on the realized location observables in the system.

\subsubsection{Defining a Simulatable EPRB Experiment} \label{sect3sub3sub1}
In many discussions of the EPRB experiment the focus is on the entangled spin states of the two particles and
the location part of the wave function is ignored. When applying the Bohm interpretation of quantum mechanics
to this system, where particle locations are the only observables with realized values in the
physical world, it is of course essential to also include these particle locations - as is done in ref.~\cite{Norsen14} 
(see also \cite{BricmontGoldsteinHemmick19a}). In the  system presented here,  particle position is
also included in the system's state space, albeit in a highly simplified manner. Moreover, and unlike in most 
discussions of EPRB experiments, the device angles are  part of the quantum system as well and will be subject 
to a last-minute change. I.e., the experiment is fully self-contained and described by a single wave function.

To achieve this,  the state space for each particle has two location-like observables in addition to spin:
a particle position observable $\hX$ and a magnet orientation observable $\hat{\Phi}$.
In order to keep the system sufficiently simple such that it can be simulated on a small computer, the particle position
and magnet orientation degrees of freedom will be restricted to a small set of discrete values.
More specifically, the state vector of each particle is
written as $\ket{\psi}=\ket{ \phi, x, \theta\pm}$. The position observable $\hX$ has six states, with (eigen)values 
$x \in \{x_r, x_a, x_{\al+},x_{\al-},x_{\bt+},x_{\bt-} \}$ 
that represent the initial `ready' state at the start of the experiment, an intermediate `all-set' state 
at which point the two spin-measuring devices (the Stern-Gerlach magnets) have been set in specific
orientations, and four `measured' states; these last four states indicate a position shift up or down 
in the $ \al$ or $\bt$ direction.
The magnet orientation observable $\hat{\Phi}$ has three states, with values $\phi \in \{0,\al,\bt\}$ that represent,
respectively,  a ready or zero-degrees state and two states that 
represent  two alternative magnet orientations. The remaining spin subspace is of course two dimensional, and has  
spin values denoted by $\theta+$ and $\theta-$; here it is sufficient to limit spin rotations to a single angle, such
that the spin can align with the angle $\phi$ of the Stern-Gerlach magnet.  As discussed above, in the eBBB 
formulation also the spin orientation angle $\theta$  is dynamic and an element of reality.

To provide quantum dynamics for the state vector $\ket{ \phi, x, \theta\pm}$,
two evolution operators are defined, $\hU^{(i)}$ and $\hU^{(f)}$,
which take the system from `ready' to `all-set' and from `all-set' to `measured' respectively.

In the first stage, the unitary evolution operator $\hU^{(i)}$ must transform the location and magnet angle 
states  from the initial `ready' state $\ket{\psi_r}=\ket{\phi_0,x_r,0\pm}$, to states 
$\ket{ \al,x_a,0\pm}$ or $ \ket{\bt,x_a,0\pm}$, 
at which point the magnets have assumed  orientation $ \al$ with probability $P_{ \al}$ or orientation 
$\bt$ with probability  $P_{\bt}=1-P_{ \al}$.
I.e., the evolution operator transforms the initial `ready' state $\ket{\psi_r}$ into an intermediate 
`all-set' state $\ket{\psi_a}$:
\be
   \ket{\psi_a} =  \hU^{(i)}\ket{\psi_r} = \hU^{(i)}\ket{\phi_0,x_r,0\pm} = 
    \gm_{\al}\ket{\al,x_a,0\pm} + \gm_{\bt}\ket{\bt,x_a,0\pm}.  \label{Eq_Psi1}
\ee
The coefficients in Eq.~(\ref{Eq_Psi1}) are such that $\vert\gm_{\al}\vert^2 = P_{\al}$
 and $\vert\gm_{\beta}\vert^2 = 1-\vert\gm_{\al}\vert^2=P_{\beta}$. 

The second stage evolution matrix then evolves this `all-set' state $\ket{\psi_a}$ into the final, 
`measured' state $\ket{\psi_m}$, which consists of a superposition of four components:
\bea
  \ket{\psi_m} = \hU^{(f)}\ket{\phi_a} & = & \gm_{\al}\bra{\al+}0\pm\rangle \ket{\al,x_{\al+},\al+} +
                                                                   \gm_{\al}\bra{\al-} 0\pm\rangle \ket{\al,x_{\al-},\al-} \\
                                  & + & \gm_{\bt} \bra{\bt+}0\pm\rangle \ket{\bt,x_{\bt+},\bt+} +
                                            \gm_{\bt} \bra{\bt-} 0\pm\rangle \ket{\bt,x_{\bt-},\bt-},  \label{Eq_Uf}
\eea
where 
\be \begin{array}{ll}
    \bra{\phi+}0+\rangle = \cos(\phi/2), & \bra{\phi+}0-\rangle = \sin(\phi/2), \\
     \bra{\phi-}0+\rangle = \sin(\phi/2),  & \bra{\phi-}0-\rangle = -\cos(\phi/2), \Sp (\phi = \al,\bt) \label{Eq_as}
\end{array} \ee
A  detailed description of the construction of these evolution matrices $\hU^{(i)}$ and $\hU^{(f)}$ can be
found in Appendix \ref{Ap_B}.

When these two evolution matrices are used in Eq.~(\ref{Eq_10}) they do not automatically lead to 
valid transition probabilities.
For this, it is required that the matrices are sufficiently close to  unity. This is not the case for the first and second stage
evolution matrix, since they produce the large transitions between the macroscopically  different `ready'  
and `all-set' state, and `all-set' and `measured' state respectively.
However, it is straightforward to introduce a sub-step evolution matrix for each stage, such
that the repeated action of this matrix reproduces the action of the full evolution matrix. For any
unitary matrix $U_{T}$ that evolves the system over a time period $T$, one can define a
sub-step matrix $U_{\eps}$ with $\eps=T/N_t$, as
\be
  U_{\eps} =U_{T}^{1/N_t}. \label{Eq_subU}
\ee
If the number of intermediate time steps $N_t$ is large enough, the sub-step matrix $U_{\eps}$ will be sufficiently
close to the unit matrix such that the consistency condition (\ref{Eq_11}) is met 
and Eq.~(\ref{Eq_10}) can be used to define transition probabilities that in
turn can be used to create stochastic trajectories for the system's location and spin values. In the simulation results
below, the time units will be such that  $T=1$ with $T$ the duration of stage one and two of the experiment.

\subsubsection{Numerical Simulation of the EPRB Spin Experiment -- Stage 1} \label{sect3sub3sub2}
To set the scene and get familiar with the type of results produced through numerical simulation of this system, 
the time evolution in the initial stage of the experiment will be discussed first. 
Here, the evolution matrix consists of three independent components for the $x$, $\phi$ and $\sg$
subsystems, $U^{(i)}= U^{(i,\phi)}U^{(i,x)}U^{(i,\sigma)}$.
For the location dynamics, it is sufficient to only consider the `ready' and `all-set' values, reducing the
evolution matrix for this subsystem to a $2\times 2$ size. 
Then the three evolution matrices can be written as 
(cf.~Eqs.~(\ref{Eq_BU1phi}, \ref{Eq_BU1x} and \ref{Eq_BU1spin})  in Appendix \ref{Ap_B})
\be
        U^{(i,\phi)}= \left( \ba{ccc} 0       &        0         &   1 \\ 
                                             \gm_{\al} & -\gm_{\bt}^* &  0 \\ 
                                             \gm_{\bt} & \gm_{\al}^*  &  0  \ea \right) ,   \label{Eq_52}
\Sp   U^{(i,x)} =  \left( \ba{cc} 0 & 1  \\
                                                   1 &  0   \ea \right),
\Sp  U^{(i,\sigma)} = \left( \ba{cc} 1 &  0 \\ 0 &  1 \ea \right).
\ee
The matrix elements are computed from Eqs.~(\ref{Eq_BU1phi}),  (\ref{Eq_BU1x})  and (\ref{Eq_BU1spin})
in Appendix \ref{Ap_B}.
The simulation results shown below, are obtained using $\gm_{\al}=\sin(\pi/5)$ and $\gm_{\bt}=\cos(\pi/5)$.

As is implied by the unit matrix for the particle spin evolution, neither the spin values, nor the best-matching 
representation  obtained when the particles are created in the `ready' state will change during the subsequent
progression in time. 
Therefore, the spin sector can be ignored and in this initial stage the focus will be on the $x$ and $\phi$ 
trajectories for one of the particles.\footnote{Even though the spin representation does not change, the best-match
approach discussed in section \ref{sect2sub3} will select a preferred representation. This will be discussed in
the section describing the stage two simulation results.}
With these simplifications, the  Schr\"odinger equation for the $\phi$-$x$ system of one of the particles takes the form,
\be
   \ket{\phi,x}^{t+1} =( U^{(i,\phi)})^{1/N_t} ( U^{(i,x)})^{1/N_t}  \ket{\phi,x}^t, \Sp \ket{\phi,x}^0= \ket{\phi_0,x_r}.
                        \label{Eq_SE}
\ee
It is then straightforward to substitute this evolving state and the sub-time step evolution matrix in Eq.~(\ref{Eq_10}) 
to compute the transition matrix for $\phi$ and $x$ values and use it to produce an ensemble of trajectories
for the magnet angle and location beables.

\begin{figure}
\begin{center}
\includegraphics[height=6.5cm]{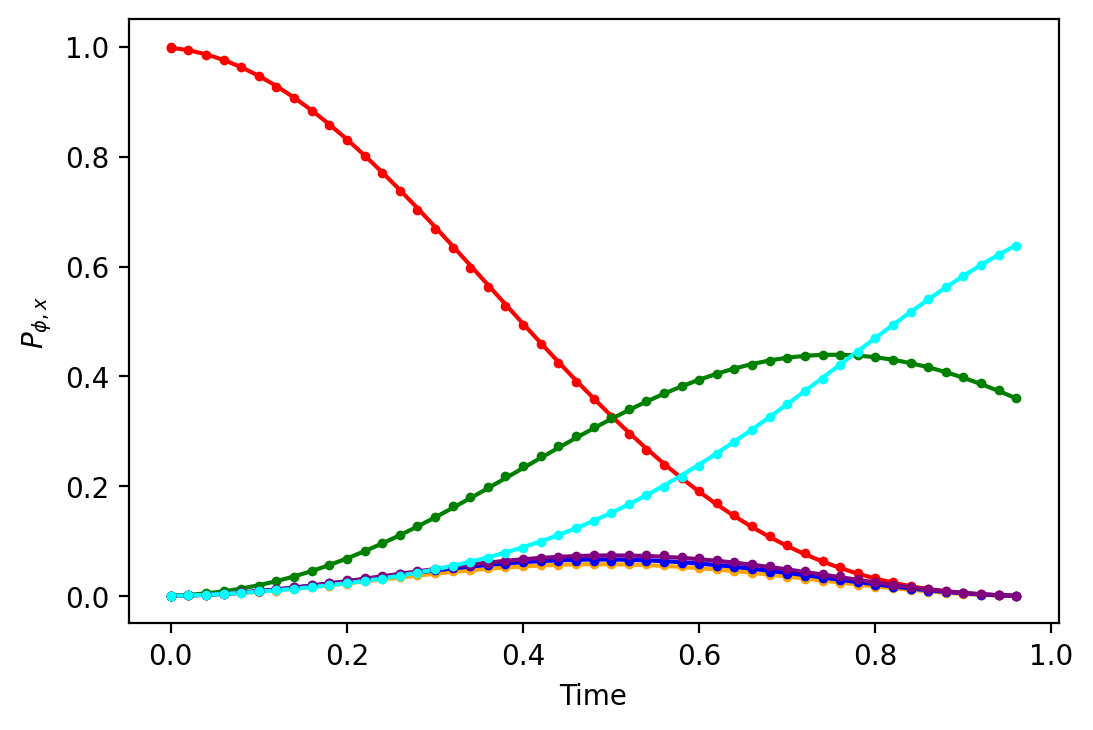}
\end{center}
\captionW{Comparison of the analytically computed probabilities (drawn lines) for the six discrete $\phi$,$x$ states 
(using Eq.~(\ref{Eq_SE})) with the numerically computed relative frequencies (discrete points with error bars
that are smaller than the markers) 
obtained from an ensemble of 50,000 trajectories computed with time step $\eps=0.02$. 
The descending curve shows the probability of the 
$\ket{ \phi_0,x_r}$ state, the two ascending curves the probabilities of the
 $\ket{ \al,x_a}$ and $ \ket{\bt,x_a}$  states. \label{Fig_St1Prob}}
\end{figure}

\begin{figure}
\begin{center}
 \includegraphics[height=3.2cm]{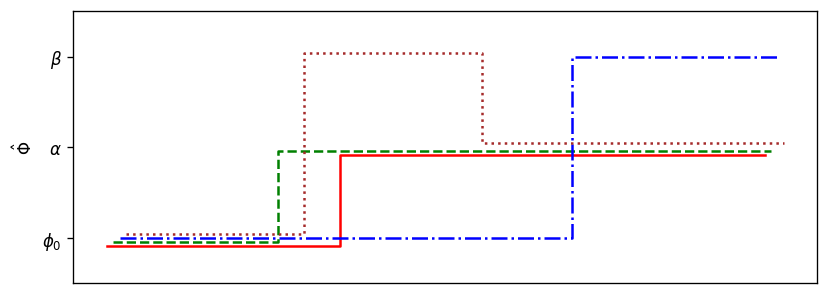}\\[-0.2cm]
\includegraphics[height=3.73cm]{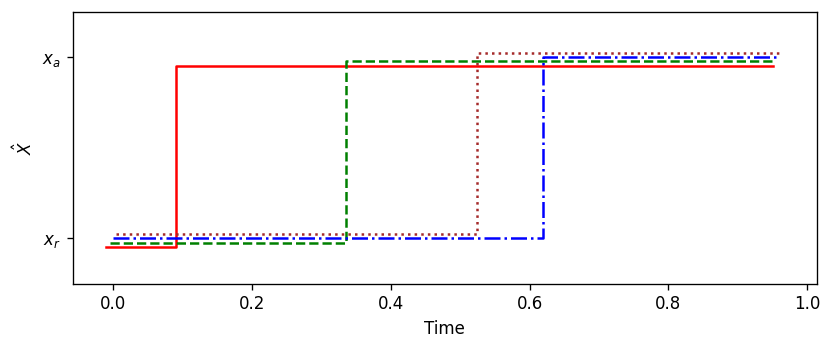}
\end{center}
\captionW{Trajectories for the discrete magnet orientation beable $\phi$ (top) and location beable $x$ (bottom). 
 All $x$-trajectories start at $x=x_r$ and end at $x=x_a$; all $\phi$-trajectories start at $\phi=\phi_0$ and
 end either at $\phi=\al$ or $\phi=\bt$. To simplify following individual trajectories, the four paths have
been slightly shifted vertically. \label{Fig_St1XF}}
\end{figure}
 
The results of such a simulation are presented in Figures \ref{Fig_St1Prob} and \ref{Fig_St1XF}.
To verify the accuracy of the discrete time dynamics, the exact probabilities 
for each of the discrete states, the $P_n$ with $n \in \{(\phi_0,x_r), (\al,x_r),(\bt,x_r),(\phi_0,x_a), 
(\al,x_a),(\bt,x_a) \}$, computed using the Schr\"odinger equation Eq.~(\ref{Eq_SE}),  are compared with 
the relative frequencies computed from an ensemble of 50,000  $\phi$ and $x$ trajectories with a time step 
size of $1/N_t = 0.02$. 
Figure \ref{Fig_St1Prob} shows that the probabilities computed from the ensemble of trajectories  reproduce
the analytical results within (the very small) statistical errors. 

Next, figure \ref{Fig_St1XF} shows four individual trajectories for magnet orientation $\phi$ (top) 
and location $x$ (bottom). 
As expected, all $x$-trajectories start at $x=x_r$ and end at $x=x_a$. The time at which the jump in value occurs is, 
of course, subject to chance, since the probability for the system to be in the $x_r$ state gradually decreases 
from $1$ at $t=0$, to $0$ at $t = 1$---as can be seen in Figure \ref{Fig_St1Prob}. 
Similarly, the trajectories for $\phi$ all start at $\phi=\phi_0$ and end at either $\phi=\al$ or $\phi=\bt$, 
with  probability $\vert\gm_{\al}\vert^2\approx 0.35$ and  $\vert\gm_{\bt}\vert^2 \approx 0.65$ respectively. 
Most trajectories have a single transition where $\phi$  jumps to either $\al$
or $\bt$. One trajectory has two transitions, first from $\phi_0$ to $\bt$, then to the end-value $\phi=\al$.
Since the evolution matrix $ U^{(i,\phi)}$ from Eq.~(\ref{Eq_52}) also has matrix elements that connect $\al$ 
and $\bt$ states, this type of trajectories, where the magnet angles can also take intermediate values that 
are different from their start or end-value, will be present in the ensemble. This is an artifact of the simple
representation of a macroscopic magnet  orientation using a single microscopic degree of freedom.

\subsubsection{Numerical Simulation of the EPRB Spin Experiment -- Stage 2} \label{sect3sub3sub2}
In a similar fashion the second stage dynamics can be simulated. Now
the evolution  is rather more involved as it  includes the conditional dynamics of the spin values in their
evolving representations.

The  evolution operator  for this stage is $\hU^{(f)}\otimes \hU^{(f)}$, with $\hU^{(f)}$ defined 
in Eq.~(\ref{Eq_Uf}) (cf.~also Eq.~(\ref{Eq_BU2}) in Appendix \ref{Ap_B}).
This operator acts on the product state of the two particles, each of which 
has a dimension equal to $N_{\phi}\times N_x \times 2 = 20$, where $N_{\phi} = 2$ is the 
dimension of the state space for the magnet orientation, $N_x = 5$ and the remaining factor
$2$ is the dimension of the spin subspace. To keep the dimension of the state space as low as possible, 
the ready state values 
$\phi=\phi_0$ and $x=x_r$  have been dropped, since they play no active part in this stage.
The combined state space therefore has dimension $400$, which is still quite manageable in a numerical simulation. 

The initial state for stage two is the `all-set' state show in Eq.~(\ref{Eq_Psi1}).
At this point,  the device angle $\phi$ for the first particle
has assumed the value $\al$ with probability $\vert\gm_{\al}\vert^2 \approx 0.35$ and value $\bt$ with probability 
$\vert\gm_{\bt}\vert^2\approx 0.65$; for particle two these probabilities are  $\vert\gm'_{\al}\vert^2 \approx 0.79$ and
$\vert\gm'_{\bt}\vert^2\approx 0.21$ respectively. Hence, the stage two initial state represents a superposition
of four different experiments in which the two device angles $(\phi_1,\phi_2)$ are $(\al,\al)$,
$(\bt,\al)$, $(\al,\bt)$ and $(\bt,\bt)$, realized with relative probabilities $0.07$, $0.14$, $0.28$ and $0.51$
respectively.
The initial location of both particles is $x=x_a$.
In this state, spin components are not correlated with $x$ or $\phi$, hence the spin state  
conditional on realized values $\phi=\al$ or $\bt$ and $x=x_a$  for the two particles is still the singlet state,
\be
   \ket{\psi^0_{\sg}} = (\ket{\theta+}\otimes\ket{\theta-} - \ket{\theta-}\otimes\ket{\theta+})/\sqrt{2}.
\ee
Following the procedure outlined in section \ref{sect2sub3}, the realized spin representations (i.e., the rotation
angles $\theta$ for the two spins) in this singlet state are found to be $\pi/2$ for both particles. 
The  spin values for both particles are, of couse, $+1/2$ or $-1/2$ with equal probability, where the spins of the two 
particles always have opposite sign.

In order to support the claim that the eBBB dynamics leads to a physically reasonable behavior of the 
elements of reality, the $\phi$, $x$, $\sg$ and $\theta$ trajectories should show the following characteristics: 
The device angles must remain at the values they assumed in the `all-set' state. 
The location values of the particles at some time
must jump from the `all-set' state to one of the `measured' states.  On trajectories with device angle 
$\al$, $x$ must jump to either $x_{\al+}$ or $x_{\al-}$ and on trajectories with device angle $\bt$, 
$x$ must jump to either $x_{\bt+}$ or $x_{\bt-}$. 
When $x$ jumps, this must be in accordance with the microscopic spin state, so also the spin representation 
angle $\theta$ must jump from $\pi/2$ to $\al$ if the device angle on this trajectory is $\al$ or to $\bt$ 
if the device angle is $\bt$, whereas the spin values of the particles must assume opposite values $\pm1/2$ 
with equal probability, but correlated in accordance with the relative angle between the two measuring devices.

\begin{figure}[ttt]
\begin{center}
\hspace{1mm}\includegraphics[height=3.cm]{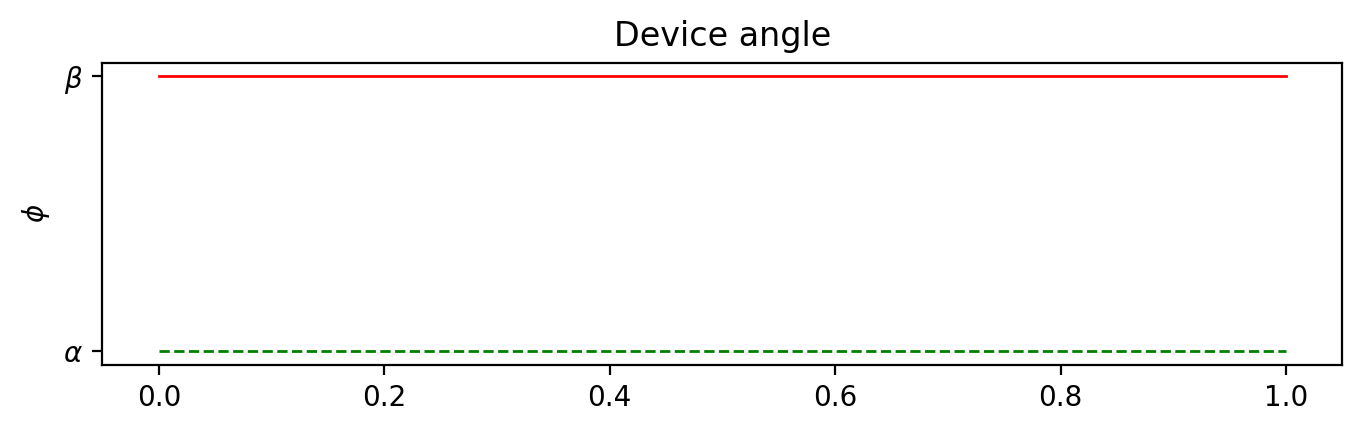}\\[-2mm]
\includegraphics[height=3.cm]{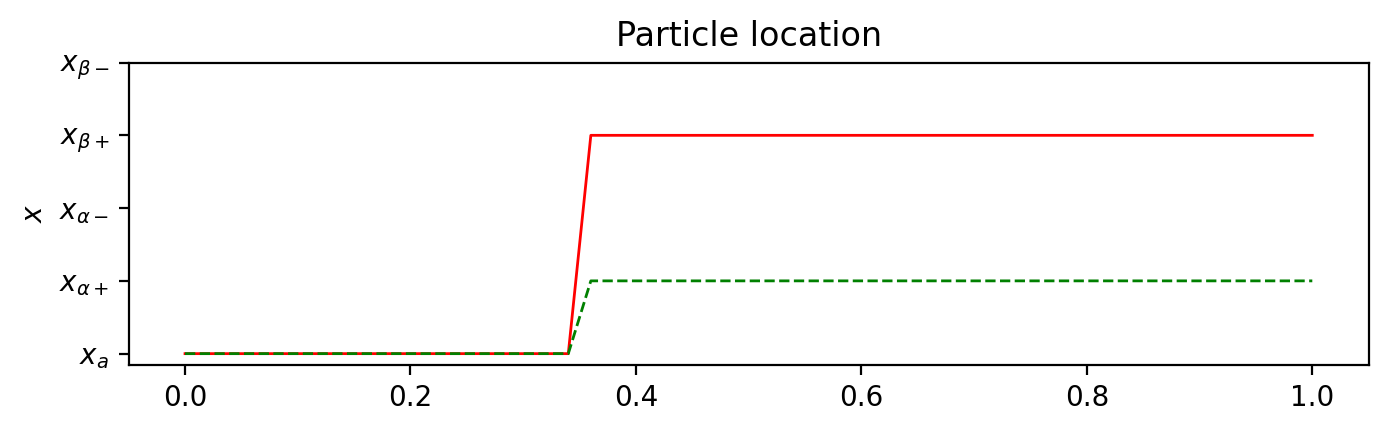}\\[-2mm]
\includegraphics[height=3.cm]{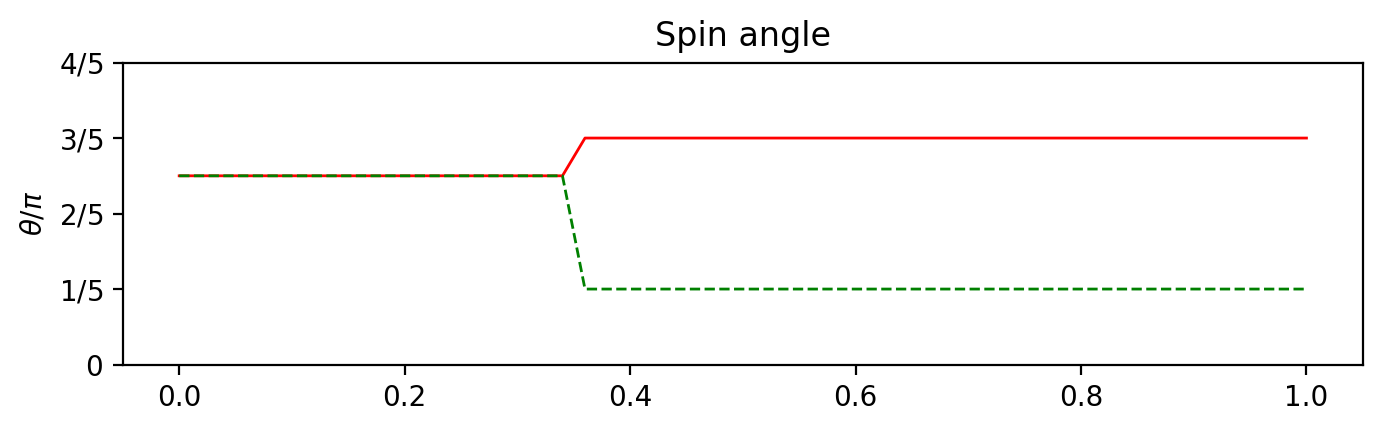}\\[-2mm]
\hspace{-1mm}\includegraphics[height=3.3cm]{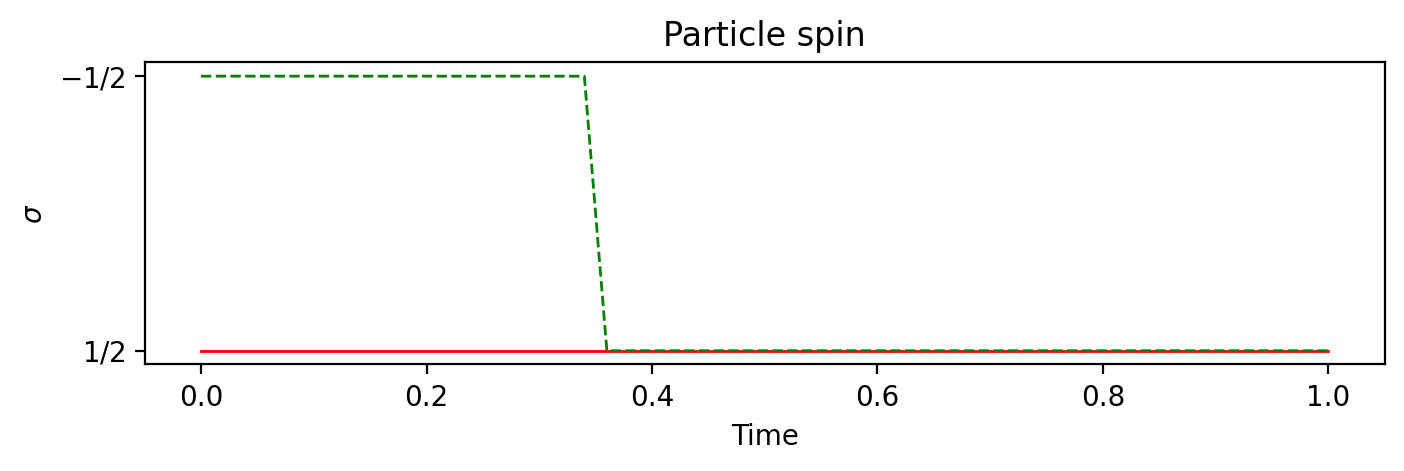}
\end{center}
\captionW{One history of the EPRB experiment simulated using the eBBB approach. The device angles in this experiment
can assume values $\al=\pi/5$ and $\bt=3\pi/5$. The top plot shows the $\phi$ trajectory for device one (solid line)
and device two (dashed line); the second plot shows the two corresponding particle location trajectories;
the third and fourth plots show  their spin representations and  spin values. The time step size $\eps=0.02$.
 \label{Fig_EPRB_Traj}}
\end{figure}

Figure \ref{Fig_EPRB_Traj} shows a set of typical trajectories from a numerical simulation of the 
EPRB experiment, which indeed confirms the behavior described above. 
In this specific history, the first device angle evolved in the first stage to $\bt$, the second
to $\al$, and both values remain unchanged in the second stage (first sub-plot).
The location variable jumps at around $t=0.35$ from $x=x_a$ to $x=x_{\bt+}$ for particle one and 
to $x=x_{\al-}$ for particle two (second sub-plot). At this same instance the spin representation angle jumps 
from $\pi/2$ to $3\pi/5=\bt$ for particle one and to $\pi/5=\al$ for particle two, 
while the spin values (which initially had opposite signs) assume the same value $+1/2$ (subplots three and four). 
A similar consistency between the realized angle, location and spin
values holds for all generated trajectories. In particular, the location and spin realizations are fully consistent
on all 50,000 generated trajectories: when $x$ jumps to $x_{\al+}$ or $x_{\al-}$ then the spin orientation
$\theta$ jumps to $\alpha$ with spin value $+1/2$ or $-1/2$ respectively, and mutatis mutandis when $x$ jumps
to  $x_{\bt+}$ or $x_{\bt-}$.

\begin{figure}[ttt]
\begin{center}
\includegraphics[height=6.5cm]{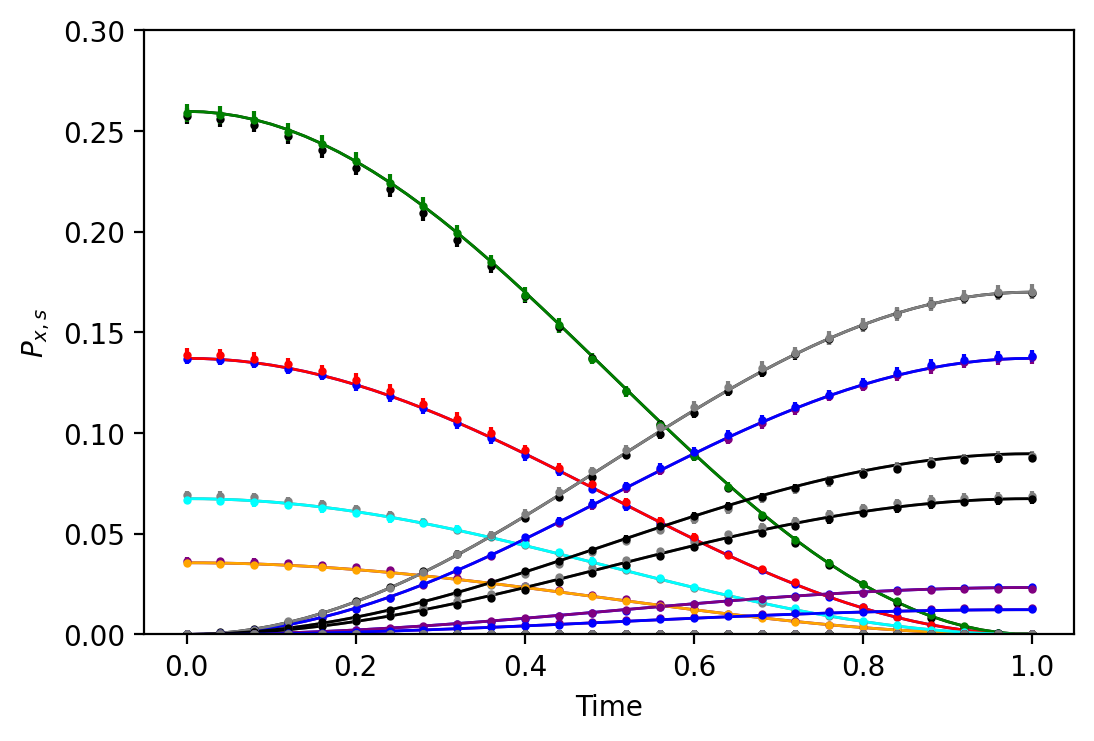}
\end{center}
\captionW{Probabilities $P_{xs}$ of the realized location ($\phi$, $x$) and spin values in the EPRB experiment.
Solid lines are computed directly from the system's wave function, dotted lines marked with error bars,
are computed from an ensemble of 50,000 trajectories. 
\label{Fig_EPRB_Pxs}}
\end{figure}

The next set of results validates the statistical accuracy of the eBBB approach. Figure \ref{Fig_EPRB_Pxs}
shows the time dependent probabilities of the combined $\phi$, $x$ and $\sg$ index values, computed from
the system's wave function, i.e., the probabilities $P^{V^{xt}t}_{xs}$ (solid lines) and the same
probabilities computed from the ensemble of trajectories (dots with error bars).
The four (degenerate pairs of) descending curves represent the initial states for the four different experiments;
the six ascending curves represent 12 final states: since $x$ and spin values are fully correlated, there are two 
(opposite spin) states for each of the two experiments with equal device angles and four states 
(with different spin combinations) for each of the two experiments with different device angles.
As expected from the discussion in section \ref{sect2sub3},
 the averages computed from the ensemble of index trajectories closely matches the result computed from
the wave function.

\begin{figure}[bbh]
\begin{center}
 \includegraphics[height=6.5cm]{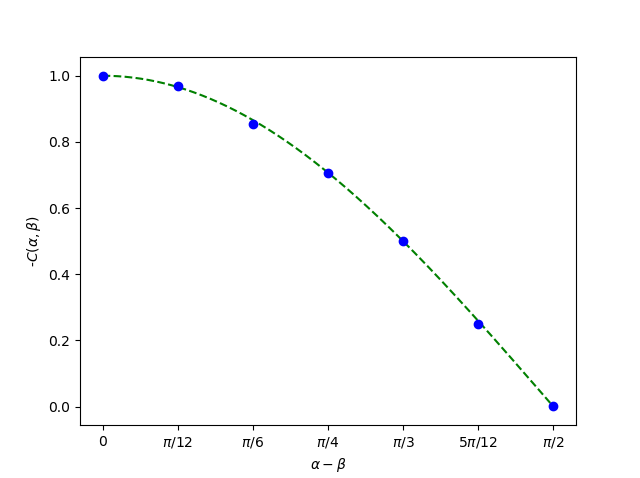}
\end{center}
\captionW{Exact (dashed line) and simulated results (dots) of (minus) the correlation between simulated 
spin values $\pm1$ of the two particles, as expressed through their $x$-values, for seven values of $\al - \bt$.
 Statistical errors are comparable with the size of the symbols. \label{Fig_corr}  }
\end{figure}

Finally, for completeness and as in an actual EPRB experiment, the  results for the average spin and 
spin-spin correlations can be computed from the ensemble of realized location variables.
Spin values and correlations can be computed using the observable $x$-values (as in a real experiment), 
which reliaby represent the not directly observable particle spins $\sg$. 
As expected the ensemble average of the spin of each particle is then
found to be zero within statistical errors, irrespective of the orientation angle $\phi$. 
The correlation, $C(\al,\bt)$ of the spin values of the two particles, measured at different values of the 
 device angles also accurately reproduces the expected result, as is illustrated in Figure \ref{Fig_corr}.

\section{Discussion} \label{sect4}
The main driver for extending the de Broglie-Bohm-Bell approach is to account for an extended 
content of the physical world where more quantum degrees of freedom beyond location (for example, spin) 
can be realized as elements of reality. 
This allows numerical simulations to more deeply probe and expose microscopic quantum behavior, 
it may be required to support the richness of our daily experiences and it supports
microscopic dynamics that is more closely in line with our common sense expectations. 
The eBBB approach explored in this paper additionally (but less fundamentally) assumes that the physical world 
is rigorously finite and discrete, which avoids complexities of dealing with infinite dimensional Hilbert spaces 
and regulates the infrared and ultraviolet divergences of quantum field theories. 
Furthermore, its stochastic dynamics can rigorously be formulated in discrete time, which then suggests a 
natural definition for a minimal time increment.

The fundamentally finite and discrete formulation
makes the eBBB formulation very ``simulatable'', which allows exploring and elucidating quantum behaviors 
such as contextuality in a very explicit and unambiguous fashion. 
Of course, the discrete space  breaks the treasured relativistic invariance of classical physics, 
and singling out a special 3-space in which the universe performs its discrete time stepping
 further breaks this invariance.  
The eBBB formulation shares this reliance on a special space-time foliation with other Bohm-type formulations 
and shares the assumption that, in extensions of this approach to relativistic quantum field theories 
\cite{BohmHileyCh11,BellCh19,DurrQFT,Vink18},  this important macroscopic invariance is restored at large scales 
where  space-time discreteness and other microscopic symmetry breaking features are hidden. 

The quantum world implied by this formulation is counterfactually definite in the generic sense that 
microscopic elements of reality exist with definite values,
irrespective of being measured or observed by humans (or other conscious beings). 
Experiments and observations are integral part of the realized world, all driven by the guidance conditions, 
the transition probabilities prescribed by the system's (or even the entire universe's) wave function.
It is contextual, but only in the specific and limited sense required to avoid the constraints of the 
Kochen-Specker theorem, 
in that only a single, compatible set of operators has values that exist as elements of reality. 
This set (or more precisely, the factorized eigenvector basis in the system's Hilbert space) 
is dynamic and evolves autonomously, also guided by the wave function and conditionally 
on the realized values of particle locations. Hence, it is  the wave function itself, and for internal
degrees of freedom the realized particle configuration, that provides the context for the beables to 
assume their appropriate values.

In a world governed by classical physics, particle dynamics is governed by locally acting forces and 
the particle's spin determines the direction in which its trajectory  deflects in an inhomogenous magnetic field.
In quantum physics, as shown explicitly in the causal Bohm interpreation \cite{Norsen14}, spin is a
contextual property of the particle and the system it interacts with, such that its dynamics is driven by a 
(non-localized) wave function. The eBBB formulation shows that it is nonetheless possible to represent this 
contextual spin as an element of reality, i.e., as a (localized) property of each individual particle.

\section*{Acknowledgments}
I would like to thank Jeffrey Barrett for valuable comments on an early draft of the paper.
 
\section*{Appendices}
\appendix
\renewcommand{\theequation}{\thesection-\arabic{equation}}
\setcounter{equation}{0}  

\section{No Surreal Trajectories with the eBBB Formulation} \label{Ap_A}
One of the non-classical features of quantum mechanics is the phenomenon that a particle can
interfere ``with itself''. In the causal Bohm interpretation this leads to trajectories with equally non-classical
features. Consider a Gaussian wave function $\psi_p(x+x_0,t)$  for a particle in one 
dimension localized around $x=-x_0$ and moving in the positive direction with momentum $p$. The
superposition 
\be
\psi(x,t_0) = (1/\sqrt{2})(\psi_p(x+x_0) + \psi_{-p}(x-x_0))  \label{Eq_A1}
\ee
then represents a system in 
which the particle at $t=t_0$ has equal probability to be localized around $x=-x_0$ moving in the positive
direction or be localized around $x=+x_0$ moving in the negative direction. After some time, when the wave
packets are in the region around $x=0$, the particle (or rather the wave function) will ``interfere with itself'', 
after which the two wave packets reemerge and continue on the paths they followed before entering the 
interference region. 

\begin{figure}[ttt]
\begin{center}
\includegraphics[height=5cm]{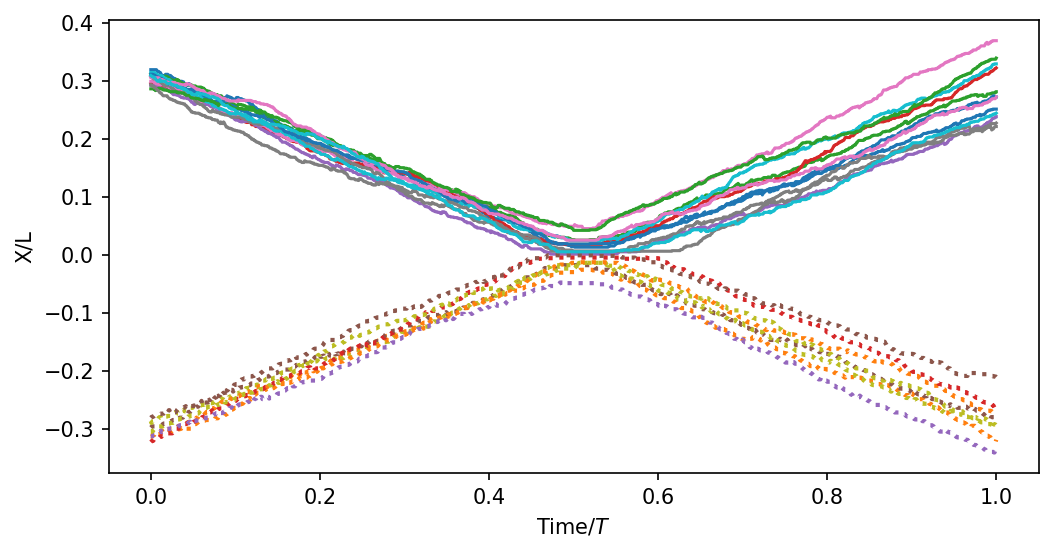}\\[-0.6cm]
\includegraphics[height=5cm]{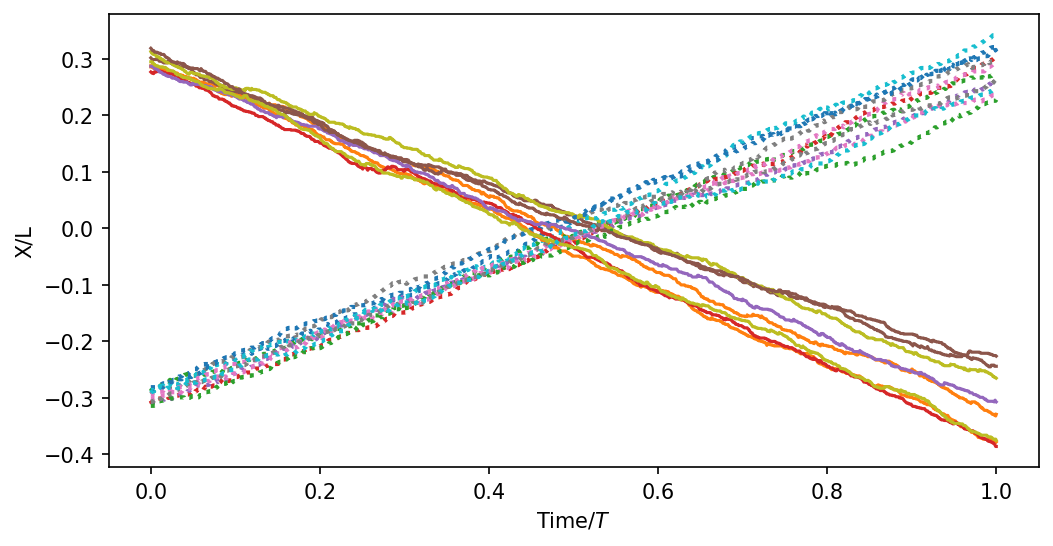}
\end{center}
\captionW{Trajectories for the system (\ref{Eq_A2}) describing a free particle with spin 1/2. 
The wave function  consists of two wave packets with opposite spin, 
which move towards each other and  are unperturbed when the packages cross around $x = 0$. 
Trajectories in the top figure (``surrealistically'') bounce back rather than cross, because they
are guided by the marginal, location only, probability current;
trajectories in the bottom figure cross $x=0$ without interference, just like the wave packets -- because they are computed with the eBBB approach of the present paper.           \label{Fig_Surreal} }
\end{figure}

In the causal Bohm interpretation (as well as in the eBBB interpretation) the trajectories
for such a particle will show the ``bouncing'' behavior discussed at length in the ``surreal trajectories" 
papers \cite{Englert93,Scully98,Durr93,Dewdney93,Hiley00,Barrett00}.
In this situation, the observation that particle trajectories will not (cannot) cross $x=0$ is perhaps surprising
at first glance, but this unexpected behavior can be justified or at least mollified by pointing to the distortions
in the wave packets, due to this non-classical self-interference, which then apparently also impacts the particle
trajectories.\footnote{More specifically, one can explain that in this interference region, the quantum potential
kicks in \cite{Bohm52,Hiley00}  and pushes the particles back from $x=0$.}
It is, however, less straightforward to argue that such a bouncing behavior of the causal Bohm trajectories is
 equally reasonable, when the particle (system) has an additional internal degree of freedom, for example spin   \cite{Dewdney93,Hiley00,Barrett00} (cf.~also ref.~\cite{Norsen14}). 
This is at the heart of the assertion in ref.~\cite{Englert93,Scully98} that these Bohmian trajectories 
should be dubbed surreal. 

The wave function (\ref{Eq_A1}) can be augmented with a binary degree of freedom. This could represent the
spin up and down state of the particle, or the two states of a resonance cavity. Then the wave function
\be
\psi(x,t_0)  = (1/\sqrt{2})(\psi_p(x+x_0)\left( \ba{c} 1 \\  0 \ea \right)
 + \psi_{-p}(x-x_0))\left( \ba{c} 0 \\  1 \ea \right)  \label{Eq_A2}
\ee
can represent a state in which one of the wave packets has triggered a particle
detection, which has flipped its spin, or flipped the binary state from $(1,0)^T$ to $(0,1)^T$ 
(as in the final stretch of the which-way experiment of ref.~\cite{Englert93}).
In the causal Bohm interpretation, also this wave function 
will  generate particle trajectories that bounce away from $x_0$, even though the wave packets now are in
orthogonal sectors of the Hilbert space and will not interfere with each other when crossing $x=0$. 
In the Bohm interpretation only particle locations are available as elements of reality - the binary state
(be it spin or resonance cavity state), which allows the wave packets to cross in configuration space, is absent. 

The easiest way to see the inevitability of the bouncing behavior of trajectories generated by the 
state (\ref{Eq_A2}), is to observe that in the Bohm approach the particle velocities $\dot{x}$ are 
obtained from the marginal probability current in which all internal particle degrees of freedom have 
been averaged over,
\be
   \dot{x}(t) = \overline{J}(x,t)/\Pb(x,t)  \label{Eq_xdot}
\ee
with
\bea
   \Pb(x,t) & = & \sum_s \psi_s^*(x,t)\psi_s(x,t) \\
      \overline{J(}x,t)  & = & (1/2mi) \sum_s \left(\psi_s^*(x,t)\del_x \psi_s(x,t) - \psi_s(x,t)\del_x \psi_s^*(x,t) \right).
\eea
For free moving particles with initial state (\ref{Eq_A2}) the marginal current is anti-symmetric in $x$ and
hence $\overline{J}=0$ at $x=0$. Together with the single valuedness of the particle velocities in (\ref{Eq_xdot})
this implies that the trajectories cannot cross $x=0$ and will show the same bouncing behavior as they do when
the guiding wave function does not have internal degrees of freedom. Since each of the two wave packets,
$\psi_1(x,t)$ and $\psi_2(x,t)$, evolves as if the other component is absent, it is difficult to accept that the
trajectories they guide along behave so differently, depending on the presence of absence of the other component.

In the eBBB formulation explored in the present paper, internal degrees of freedom (like spin) {\it are} elements
of reality and the particle trajectories also carry this degree of freedom. Therefore, the (stochastic) eBBB-trajectories
can (and will) cross each other at $x=0$, because the left and right moving particles have different spin 
(or cavity state) values. This is illustrated in Figure \ref{Fig_Surreal},
where the top figure shows a set of trajectories guided by the marginal current and the bottom 
figure trajectories ($x$-values only) guided by the transition probabilities computed using the eBBB approach.
Here, the up-  and down-moving trajectories have a different internal state index, and hence they can cross.

\section{Evolution Operators for the Simplified EPRB Experiment} \label{Ap_B}
This appendix provides further details of the simplified EPRB model used for the numerical simulation discussed
in this paper. In particular, the construction of the two evolution operators, $\hU^{(i)}$ and $\hU^{(f)}$ will
be described in more detail.

In the first stage of the experiment, the unitary evolution operator $\hU^{(i)}$ must transform the location 
and magnet angle states  from the initial state $\ket{\phi_0,x_r,0\pm}$, to states $\ket{ \al,x_a,0\pm}$ or 
$ \ket{\bt,x_a,0\pm}$, at which point the magnets have assumed  orientation $ \al$ with probability $P_{ \al}$ 
or orientation $\bt$ with probability  $P_{\bt}=1-P_{ \al}$.
I.e., the magnet orientation subspace evolution operator is defined on the three-dimensional space
spanned by $\{\ket{\phi_0},\ket{\al},\ket{\bt}\}$ and acts on the `ready' state $\ket{\phi_0}$ as
\be
   \hU^{(i,\phi)}\ket{\phi_0} = \gm_{\al}\ket{\al} + \gm_{\bt}\ket{\bt}, \label{Eq_B31}
\ee
which can be achieved, for example, by a unitary evolution operator of the form
\be
    \hU^{(i,\phi)} =   \gm_{\al}\ket{\al}\bra{\phi_0} 
                              + \gm_{\bt}\ket{\bt}\bra{\phi_0}
                                - \gm_{\bt}^*\ket{\al}\bra{\al} 
                               + \gm_{\al}^*\ket{\bt}\bra{\al}
                               +                   \ket{\phi_0}\bra{\bt}.                                   \label{Eq_BU1phi}
\ee
The complex-valued coefficients in Eq.~(\ref{Eq_B31}) are such that $\vert\gm_{\al}\vert^2 = P_{\al}$ 
and $\vert\gm_{\beta}\vert^2 = 1-\vert\gm_{\al}\vert^2=P_{\beta}$. 
 The location subspace evolution operator for the initial stage acts as
\be
   \hU^{(i,x)}\ket{x_r} = \ket{x_a},
\ee
which implies that it can, for example,  be defined as
\be
  \hU^{(i,x)} = \sum_{i\ne a,r} \ket{x_i}\bra{x_i} + \ket{x_r}\bra{x_a} + \ket{x_a}\bra{x_r}.  \label{Eq_BU1x}
\ee
In this initial stage of the experiment the spin state does not change, hence the identity matrix can be used for its evolution:
\be
 \hU^{(i,\sigma)} = \ket{0+}\bra{0+} +  \ket{0-}\bra{0-}.    \label{Eq_BU1spin}
\ee
Since the three substates evolve independently,  the combined evolution operator for the first stage is the
direct product
\be
   \hU^{(i)} =   \hU^{(i,\phi)} \otimes \hU^{(i,x)} \otimes \hU^{(i,\sigma)},   \label{Eq_BU1}
\ee
and the  `all-set' state is generated from the `ready' state as,
\be
 \ket{\psi_a} = \hU^{(i)}\ket{\phi_0,x_r,0\pm} = \gm_{\al}\ket{\al,x_a,0\pm} + \gm_{\bt}\ket{\bt,x_a,0\pm}. \label{Eq_BPsi1}
\ee

In the second and final stage of the experiment, the interaction between Stern-Gerlach magnet and particle spin 
leads to a combined evolution in which the $\ket{\al,x_a,0\pm}$ state transforms into $\ket{\al,x_{\al\pm},\al\pm}$
and the $\ket{\bt,x_a,0\pm}$ state transforms into $\ket{\bt,x_{\bt\pm},\bt\pm}$.
I.e., the $\al$ substate from the r.h.s.~in Eq.~(\ref{Eq_BPsi1}) evolves into a state with spin in the $\al+$ or $\al-$
direction and corresponding location value $x_{\al+}$ or $x_{\al-}$ and similarly for the $\bt$ substate. 
In this stage, unlike in the first, there is no change in the device angles, hence the
evolution operator can be trivial; likewise, there is no force acting on the particle spin, other than the impact
of the Stern-Gerlach device, which forces the spin axis to be aligned with the magnet orientation.
Therefore, in this second stage the realized macroscopic magnet orientation and changes in particle location and 
spin orientation become correlated. This evolution can no longer be described with a factorized operator of the 
form shown in Eq.~(\ref{Eq_BU1}). 
However, introducing projection operators allows employing suitable combinations of subspace 
evolution operators that implement the correlated state transitions. 

Since there are four different components in the final state of each
particle---reflecting the two values of the magnet angle and for each angle the two spin values---the combined evolution
operator is composed of four terms:
\bea
   \hU^{(f)} & = &  \hP^{(\phi)}_{\al} \otimes \hU^{(x)}_{ x_{\al+}} \otimes \hP^{(\sg)}_{\al+}
                       +\hP^{(\phi)}_{\al} \otimes  \hU^{(x)}_{ x_{\al-}} \otimes \hP^{(\sg)}_{\al-} \\[1mm]
               & + & \hP^{(\phi)}_{\bt} \otimes  \hU^{(x)}_{ x_{\bt+}} \otimes  \hP^{(\sg)}_{\bt+}
                       + \hP^{(\phi)}_{\bt} \otimes  \hU^{(x)}_{ x_{\bt-}} \otimes \hP^{(\sg)}_{\bt-}.    \label{Eq_BU2}
\eea
The projection operators on the magnet angle subspaces in this expression are defined as
\be
    \hP^{(\phi)}_{\al} = \ket{\al}\bra{\al}, \Sp \hP^{(\phi)}_{\bt} = \ket{\bt}\bra{\bt},
\ee
the $x$-space evolution operators map the $x_a$ state to the state indicated by their subscripts and can be defined as
\be
\hU^{(x)}_{ x_k} = \sum_{i\ne a,k} \ket{x_i}\bra{x_i} + \ket{x_k}\bra{x_a} + \ket{x_a}\bra{x_k},\Sp {\rm for}\;  
                            k=\al+,\al-,\bt+,\bt-,
\ee
and the projectors on the (rotated) positive and negative spin states are
\be
   \hP^{(\sg)}_{\al+}  =  \ket{\al+}\bra{\al+}, \Sp  \hP^{(\sg)}_{\al-}  =  \ket{\al-}\bra{\al-}. \label{Eq_BPphi}
\ee
A similar definition applies to the projectors on spin states in the $\bt$ direction.
Since the location subspace evolution matrices are unitary and the projectors are orthogonal, one can readily
verify that also the combined evolution matrix $\hU^{(f)}$ is unitary.

It is now straightforward to show that the combined evolution operator for the two particles
\be
   U^{(f)}U^{(i)}\otimes U^{(f)}U^{(i)},   \label{Eq_BUtot}
\ee
 transforms the initial spin-singlet `all-set' state
\be
   \ket{\psi_r} = \left( \ket{\phi_0,x_r,0+}\otimes\ket{\phi_0,x_r,0-} - 
                                                    \ket{\phi_0,x_r,0-}\otimes\ket{\phi_0,x_r,0+} \right)/\sqrt{2} ,
\ee
to an end state with the expected correlations between measured  values of the  spins of the two particles. 
To simplify notations, from now on the $\otimes$ symbol will only be used to indicate direct products between 
the states or operators for particle one and two.

After applying the initial-stage evolution operator, the `ready' state turns into the following `all-set' state,
\bea
   \ket{\psi_a} & = & U^{(i)}\otimes U^{(i)}\ket{\psi_r} =   \\[1mm]
                    &\Sp&  \left( (\gm_{\al}\ket{\al,x_a,0+} + \gm_{\bt}\ket{\bt,x_a,0+}) 
                            \otimes (\gm'_{\al}\ket{\al,x_a,0-} + \gm'_{\bt}\ket{\bt,x_a,0-})\right. \\[1mm]
                 &  -  &  \left. \; (\gm_{\al}\ket{\al,x_a,0-}  + \gm_{\bt}\ket{\bt,x_a,0-}) 
                           \otimes (\gm'_{\al}\ket{\al,x_a,0+} + \gm'_{\bt}\ket{\bt,x_a,0+})
 \right)/\sqrt{2}.   \label{Eq_BAllSet}
\eea
The same evolution operators are used for particle one and two; i.e., for both particles, the same two orientation 
angles $\al$ and $\bt$ for the Stern-Gerlach magnet can be assumed, only the coefficients $\gm_{\al,\bt}$ 
that determine the relative probability to realize these orientations can be different, as is indicated by the 
prime on these coefficients in the `all-set' state of particle two.

The second stage evolution operator will produce a `measured' state which is a superposition of states for the four
different combinations of the magnet orientations:
\bea
   \ket{\psi_m} & = &  U^{(f)}\otimes U^{(f)}\ket{\psi_a} =   \\[1mm]
                   &\Sp&       \gm_{\al}\gm'_{\al} (\ket{\al}\otimes \ket{\al})\ket{\psi_{\al,\al} }
                               + \gm_{\al}\gm'_{\bt} (\ket{\al}\otimes \ket{\bt})\ket{\psi_{\al,\bt} } \\[1mm]
                        & + & \gm_{\bt}\gm'_{\al} (\ket{\bt}\otimes \ket{\al})\ket{\psi_{\bt,\al} }
                              + \gm_{\bt}\gm'_{\bt} (\ket{\bt}\otimes \ket{\bt})\ket{\psi_{\bt,\bt} }.
\eea
This shows that the full state is a linear combination of four\footnote{The three states $\ket{\psi_{\al,\al}}$,
$\ket{\psi_{\bt,\al}}$ and $\ket{\psi_{\bt,\bt}}$ that are not shown, follow by suitably swapping $\al$ and 
$\bt$ indices.} conditional states
$\ket{\psi_{\al,\bt}}\equiv(U^{(f)}\otimes U^{(f)})(\ket{x_a,\al+}\otimes\ket{x_a,\bt-} - \ket{x_a,\al-}\otimes\ket{x_a,\bt+})$, which can be computed as
\bea
 \ket{\psi_{\al,\bt}}  & = & \left(  (c_{\al}\ket{x_{\al+},\al+} + s_{\al} \ket{x_{\al-},\al-})
                                     \otimes (s_{\bt}\ket{x_{\bt+},\bt+} - c_{\bt} \ket{x_{\bt-},\bt-})  \right.\\[1mm]
                                 &&-  \left. (s_{\al}\ket{x_{\al+},\al+}  - c_{\al} \ket{x_{\al-},\al-})
                                    \otimes (c_{\bt}\ket{x_{\bt+},\bt+} +s_{\bt} \ket{x_{\bt-},\bt-})  \right) / \sqrt{2}\\[1mm]
     & = & \left(  -s_{\al-\bt} \ket{x_{\al+},\al+}\otimes \ket{x_{\bt+},\bt+} - c_{\al-\bt}\ket{x_{\al+},\al+}
                                   \otimes \ket{x_{\bt-},\bt-}\right.  \\[1mm]
       &&\left. +  c_{\al-\bt} \ket{x_{\al-},\al-}  \otimes \ket{x_{\bt+},\bt+} -  s_{\al-\bt} \ket{x_{\al-},\al-}  
                                  \otimes \ket{x_{\bt-},\bt-} \right)/ \sqrt{2}.    \label{Eq_BPsiab}
\eea
To obtain these equalities, the following results were used for the projections of the 
$\ket{0\pm}$ states on the rotated $\ket{\al\pm}$ states:
\be
   \bra{\al+}0+\rangle = c_{\al}, \Sp \bra{\al+}0-\rangle = \bra{\al-}0+\rangle = s_{\al}, \Sp
                              \bra{\al-}0-\rangle = -c_{\al};  \label{Eq_Bas}
\ee
with the following abbreviations for the sines and cosines in the expressions above:
\be
  c_{\phi} \equiv \cos(\phi/2), \Sp s_{\phi} \equiv \sin(\phi/2), \Sp \phi \in\{ \al, \bt, \al-\bt \}.
\ee

\end{document}